\documentstyle[preprint,aps,epsf,floats]{revtex} 
\begin{document}

\preprint{\tighten \vbox{\hbox{CALT-68-2141}  \hbox{UCSD/PTH 97-34} 
  \hbox{hep-ph/9711248} \hbox{} }}

\title{Comment on $V_{ub}$ from Exclusive Semileptonic $B$ and $D$ Decays}
 
\author{Zoltan Ligeti$\,^a$, Iain W.\ Stewart$\,^b$ and Mark B.\ Wise$\,^b$ }

\address{
  $^a$Department of Physics, University of California San Diego, 
    La Jolla, CA 92093 \\
  $^b$California Institute of Technology, Pasadena, CA 91125  }

\maketitle

{\tighten
\begin{abstract}%
The prospects for determining $|V_{ub}|$ from exclusive $B$ semileptonic decay
are discussed.  The double ratio of form factors $(f^{(B\to\rho)}/f^{(B\to
K^*)})/$ $(f^{(D\to\rho)}/f^{(D\to K^*)})$ is calculated using chiral
perturbation theory.  Its deviation from unity due to contributions that are
non-analytic in the symmetry breaking parameters is very small.  Combining
experimental data obtainable from $B\to\rho\,\ell\,\bar\nu_\ell$, $B\to
K^*\ell\,\bar\ell$ and $D\to\rho\,\bar\ell\,\nu_\ell$ can lead to a model 
independent determination of $|V_{ub}|$ with an uncertainty from theory of 
about 10\%.

\end{abstract}
}

\newpage

The next generation of $B$ decay experiments will test the flavor sector of the
standard model at high precision.  The basic approach is to determine the
elements of the CKM matrix using different methods and then check for the
consistency of these results.  At the present time $CP$ non-conservation has
only been observed in kaon decay arising from $K^0-\bar K^0$ mixing.  Many
extensions of the minimal standard model (e.g.,~models with several Higgs
doublets or low energy supersymmetry) have new particles with weak scale masses
that contribute to flavor changing neutral current processes like $K^0-\bar
K^0$ mixing, $B^0-\bar B^0$ mixing, $B\to K^* \gamma$, etc., at a level
comparable to the standard model.

At the present time, the magnitude of the $b \to u$ CKM matrix element is
determined by comparing experimental results on the inclusive electron spectrum
in the endpoint region with phenomenological models \cite{incl}, or by
comparing experimental results on $B\to\rho\,\ell\,\bar\nu_\ell$ and
$B\to\pi\,\ell\,\bar\nu_\ell$ with phenomenological models and lattice QCD
results \cite{excl}.  These two approaches yield remarkably consistent
determinations of $|V_{ub}|$, but have large uncertainties.

This paper is a comment on the proposal to determine $|V_{ub}|$ \cite{IsWi,lw}
using a combination of heavy quark symmetry \cite{HQS} and $SU(3)$ flavor
symmetry.  The basic idea is to compare $D\to K^*\,\bar\ell\,\nu_\ell$ with the
Cabibbo suppressed decay $D\to\rho\,\bar\ell\,\nu_\ell$.  Using heavy quark
symmetry the $SU(3)$ violations in the form factors that occur in these decays
are related to those that occur in a comparison of $B\to K^*\ell\,\bar\ell$ (or
$B\to K^*\,\nu_\ell\,\bar\nu_\ell$) with $B \to \rho\,\ell\,\bar\nu_\ell$. 
Therefore, experimental data on $B\to K^*\ell\,\bar\ell$ in conjunction with
data on $D\to\rho\,\bar\ell\,\nu_\ell$ and $D\to K^*\,\bar\ell\,\nu_\ell$ can
be used to determine $|V_{ub}|$.  This proposal is complementary to other
approaches for determining $|V_{ub}|$, since it relies on the standard model
correctly describing the rare flavor changing neutral current process $B\to
K^*\ell\,\bar\ell$.  

In this letter we compute corrections to these form factor relations which
violate both chiral and heavy quark symmetry, and are non-analytic in the
symmetry breaking parameters.  We also reconsider the influence of long
distance effects on the extraction of the $B\to K^*$ form factors from $B\to
K^*\ell\,\bar\ell$.

We denote the form factors relevant for semileptonic transitions
between a pseudoscalar meson $P^{(Q)}$, containing a heavy quark $Q$, and a
member of the lowest lying multiplet of vector mesons, $V$, by $g^{(H\to V)}$,
$f^{(H\to V)}$ and $a_\pm^{(H\to V)}$, where
\begin{eqnarray}\label{ffdef}
\langle V(p',\epsilon) |\,\bar q\,\gamma_\mu\, Q\,| H(p)\rangle
&=& i\,g^{(H\to V)}\, \varepsilon_{\mu\nu\lambda\sigma}\, \epsilon^{*\nu}\,
  (p+p')^\lambda\, (p-p')^\sigma \,, \\*
\langle V(p',\epsilon) |\,\bar q\,\gamma_\mu\gamma_5\, Q\,| H(p)\rangle
&=& f^{(H\to V)}\,\epsilon^*_\mu 
  + a_+^{(H\to V)}\,(\epsilon^*\cdot p)\,(p+p')_\mu 
  + a_-^{(H\to V)}\,(\epsilon^*\cdot p)\,(p-p')_\mu \nonumber\,,
\end{eqnarray}
and $\varepsilon^{0123}=-\varepsilon_{0123}=1$.  We view the form factors $g$,
$f$ and $a_\pm$ as functions of the dimensionless variable $y=v\cdot v'$, where
$p=m_H\,v$, $p'=m_V\,v'$, and $q^2=(p-p')^2=m_H^2+m_V^2-2m_H\,m_V\,y$. 
(Although we are using the variable $v\cdot v'$, we are not treating the quarks
in $V$ as heavy.)  The experimental values for the $D\to
K^*\,\bar\ell\,\nu_\ell$ form factors assuming nearest pole dominance for the
$q^2$ dependences are \cite{E791b}
\begin{eqnarray}\label{ffexp}
f^{(D\to K^*)}(y) &=& {(1.9\pm0.1)\,{\rm GeV}\over 1+0.63\,(y-1)}\,, 
  \nonumber\\*
a_+^{(D\to K^*)}(y) &=& -{(0.18\pm0.03)\,{\rm GeV}^{-1}\over 1+0.63\,(y-1)}\,, 
  \nonumber\\*
g^{(D\to K^*)}(y) &=& -{(0.49\pm0.04)\,{\rm GeV}^{-1}\over 1+0.96\,(y-1)}\,.
\end{eqnarray}
The shapes of these form factors are beginning to be probed experimentally
\cite{E791b}.  The form factor $a_-$ is not measured because its contribution
to the $D\to K^*\,\bar\ell\,\nu_\ell$ decay amplitude is suppressed by the
lepton mass.  The minimal value of $y$ is unity (corresponding to the zero
recoil point) and the maximum value of $y$ is
$(m_D^2+m_{K^*}^2)/(2m_D\,m_{K^*})\simeq1.3$ (corresponding to $q^2=0$).  Note
that $f(y)$ changes by less than 20\% over the whole kinematic range $1<y<1.3$.
In the following analysis we will extrapolate the measured form factors to the
larger region $1<y<1.5$.  The full kinematic region for
$B\to\rho\,\ell\,\bar\nu_\ell$ is $1<y<3.5$.

The differential decay rate for semileptonic $B$ decay (neglecting the lepton
mass, and not summing over the lepton type $\ell$) is
\begin{equation}\label{SLrate}
{{\rm d}\Gamma(B\to\rho\,\ell\,\bar\nu_\ell)\over{\rm d}y} 
  = {G_F^2\,|V_{ub}|^2\over48\,\pi^3}\, m_B\, m_\rho^2\, S^{(B\to\rho)}(y) \,.
\end{equation}
Here $S^{(H\to V)}(y)$ is the function
\begin{eqnarray}\label{shape}
S^{(H\to V)}(y) &=& \sqrt{y^2-1}\, \bigg[ \Big|f^{(H\to V)}(y)\Big|^2\,
  (2+y^2-6yr+3r^2) \nonumber\\*
&&\phantom{} + 4{\rm Re} \Big[a_+^{(H\to V)}(y)\, f^{(H\to V)}(y)\Big]
  m_H^2\, r\, (y-r) (y^2-1) \nonumber\\*
&&\phantom{} + 4\Big|a_+^{(H\to V)}(y)\Big|^2 m_H^4\, r^2 (y^2-1)^2 + 
  8\Big|g^{(H\to V)}(y)\Big|^2 m_H^4\, r^2 (1+r^2-2yr)(y^2-1)\, \bigg] 
  \nonumber\\*
&=& \sqrt{y^2-1}\, \Big|f^{(H\to V)}(y)\Big|^2\, (2+y^2-6yr+3r^2)\, 
  [1+\delta^{(H\to V)}(y)] \,,
\end{eqnarray}
with $r=m_V/m_H$.  The function $\delta^{(H\to V)}$ depends on the ratios of
form factors $a_+^{(H\to V)}/f^{(H\to V)}$ and $g^{(H\to V)}/f^{(H\to V)}$. 
$S^{(B\to\rho)}(y)$ can be estimated using combinations of $SU(3)$ flavor
symmetry and heavy quark symmetry.  $SU(3)$ symmetry implies that the $\bar
B^0\to\rho^+$ form factors are equal to the $B\to K^*$ form factors and the
$B^-\to\rho^0$ form factors are equal to $1/\sqrt2$ times the $B\to K^*$ form
factors.  Heavy quark symmetry implies the relations \cite{IsWi}
\begin{eqnarray}\label{BDrel}
f^{(B\to K^*)}(y) &=& \left({m_B\over m_D}\right)^{1/2} 
  \bigg[{\alpha_s(m_b)\over \alpha_s(m_c)}\bigg]^{-6/25}\, 
  f^{(D\to K^*)}(y)\,, \nonumber\\*
a_+^{(B\to K^*)}(y) &=& \left({m_D\over m_B}\right)^{1/2} 
  \bigg[{\alpha_s(m_b)\over \alpha_s(m_c)}\bigg]^{-6/25}\,
  a_+^{(D\to K^*)}(y) \,, \nonumber\\*
g^{(B\to K^*)}(y) &=& \left({m_D\over m_B}\right)^{1/2} 
  \bigg[{\alpha_s(m_b)\over \alpha_s(m_c)}\bigg]^{-6/25}\, 
  g^{(D\to K^*)}(y)\,.
\end{eqnarray}
The second relation is obtained using $a_-^{(D\to K^*)}=-a_+^{(D\to K^*)}$, 
valid in the large $m_c$ limit.

Using Eq.~(\ref{BDrel}) and $SU(3)$ to get $\bar
B^0\to\rho^+\,\ell\,\bar\nu_\ell$ form factors (in the region $1<y<1.5$) from
those for $D\to K^*\bar\ell\,\nu_\ell$ given in Eq.~(\ref{ffexp}) yields
$S^{(B\to\rho)}(y)$ plotted in Fig.~1 of Ref.~\cite{lw}.  The numerical values
in Eq.~(\ref{ffexp}) differ slightly from those used in Ref.~\cite{lw}.  This
makes only a small difference in $S^{(B\to\rho)}$, but changes
$\delta^{(B\to\rho)}$ more significantly.  In the region $1<y<1.5$,
$|\delta^{(B\to\rho)}(y)|$ defined in Eq.~(\ref{shape}) is less than 0.06,
indicating that $a_+^{(B\to\rho)}$ and $g^{(B\to\rho)}$ make a small
contribution to the differential rate in this region.

This prediction for $S^{(B\to\rho)}$ can be used to determine $|V_{ub}|$ from a
measurement of the $B\to\rho\,\ell\,\bar\nu_\ell$ semileptonic decay rate in
the region $1<y<1.5$.  This method is model independent, but cannot be expected
to yield a very accurate value of $|V_{ub}|$.  Typical $SU(3)$ violations are
at the $10-20$\% level and one expects similar violations of heavy quark
symmetry.  

Ref.~\cite{lw} proposed a method for getting a value of $S^{(B\to\rho)}(y)$
with small theoretical uncertainty.  They noted that the ``Grinstein-type"
\cite{Gtdr} double ratio
\begin{equation}\label{Gtdr}
R(y) = \Big[ f^{(B\to\rho)}(y) / f^{(B\to K^*)}(y) \Big] \Big/
  \Big[ f^{(D\to\rho)}(y) / f^{(D\to K^*)}(y) \Big]
\end{equation}
is unity in the limit of $SU(3)$ symmetry or in the limit of heavy quark
symmetry.  Corrections to the prediction $R(y)=1$ are suppressed by
$m_s/m_{c,b}$ ($m_{u,d} \ll m_s$) instead of $m_s/\Lambda_{\rm QCD}$ or
$\Lambda_{\rm QCD}/m_{c,b}$.  Since $R(y)$ is very close to unity, the 
relation 
\begin{equation}\label{magic}
S^{(B\to\rho)}(y) = S^{(B\to K^*)}(y)\,
  \left|{f^{(D\to\rho)}(y)\over f^{(D\to K^*)}(y)}\right|^2\,
  \bigg({m_B-m_\rho\over m_B-m_{K^*}}\bigg)^2\,,
\end{equation}
together with measurements of $|f^{(D\to K^*)}|$, $|f^{(D\to\rho)}|$, and
$S^{(B\to K^*)}$ will determine $S^{(B\to\rho)}$ with small theoretical
uncertainty.  The last term on the right-hand-side makes Eq.~(\ref{magic})
equivalent to Eq.~(\ref{Gtdr}) in the $y\to1$ limit.  The ratio of the
$(2+y^2-6yr+3r^2)\,[1+\delta^{(B\to V)}(y)]$ terms makes only a small and
almost $y$-independent contribution to $S^{(B\to\rho)}/S^{(B\to K^*)}$ in the
range $1<y<1.5$.  Therefore, corrections to Eq.~(\ref{magic}) are at most a few
percent larger than those to $R(y)=1$.

$|f^{(D\to K^*)}|$ has already been determined.  $|f^{(D\to\rho)}|$ may be
obtainable in the future, for example from experiments at $B$ factories, where
improvements in particle identification help reduce the background from the
Cabibbo allowed decay.  The measurement ${\cal
B}(D\to\rho^0\,\bar\ell\,\nu_\ell)/{\cal B}(D\to\bar
K^{*0}\,\bar\ell\,\nu_\ell) = 0.047\pm0.013$ \cite{E791a} already suggests that
$|f^{(D\to\rho)}/f^{(D\to K^*)}|$ is close to unity.  Assuming $SU(3)$ symmetry
for the form factors, but keeping the explicit $m_V$-dependence in $S^{(D\to
V)}(y)$ and in the limits of the $y$ integration, the measured form factors in
Eq.~(\ref{ffexp}) imply ${\cal B}(D\to\rho^0\,\bar\ell\,\nu_\ell)/{\cal
B}(D\to\bar K^{*0}\,\bar\ell\,\nu_\ell) = 0.044$.\footnote{This prediction
would be $|V_{cd}/V_{cs}|^2/2\simeq0.026$ with $m_\rho=m_{K^*}$.  Phase space
enhances $D\to\rho$ compared to $D\to K^*$ to yield the quoted prediction.}
$S^{(B\to K^*)}$ is obtainable from experimental data on $B\to
K^*\,\nu_\ell\,\bar\nu_\ell$ or $B\to K^*\ell\,\bar\ell$.  While the former
process is very clean theoretically, it is very difficult experimentally.  A
more realistic goal is to use $B\to K^*\ell\,\bar\ell$, since CDF expects to
observe $400-1100$ events in the Tevatron run II (if the branching ratio is in
the standard model range) \cite{CDF2}.  There are some uncertainties associated
with long distance nonperturbative strong interaction physics in this
extraction of $S^{(B\to K^*)}(y)$.  To use the kinematic region $1<y<1.5$, the
form factor ratio $f^{(D\to\rho)}/f^{(D\to K^*)}$ in Eq.~(\ref{magic}) must be
extrapolated to a greater region than what can be probed experimentally.  For
this ratio, the uncertainty related to this extrapolation is likely to be
small.

The main purpose of this comment is to examine the deviation of $R$ from unity
using chiral perturbation theory.  We find that it is at the few percent level.
The uncertainty from long distance physics in the extraction of $S^{(B\to
K^*)}$ is also reviewed.  On average, in the region $1<y<1.5$, this is probably
less than a 10\% effect on the $B\to K^*\ell\,\bar\ell$ decay rate. 
Consequently, a determination of $|V_{ub}|$ from experimental data on $D\to
K^*\bar\ell\,\nu_\ell$, $D\to\rho\,\bar\ell\,\nu_\ell$, $B\to
K^*\ell\,\bar\ell$ and $B\to\rho\,\ell\,\bar\nu_\ell$ with an uncertainty from
theory of about 10\% is feasible.

\medskip\noindent
{\bf (i) Chiral perturbation theory for $R$}
\medskip

The leading deviation of $R$ from unity can be calculated using a combination
of heavy hadron chiral perturbation theory for the mesons containing a heavy
quark and for the lowest lying vector mesons.  We adopt the notations and
conventions of Refs.~\cite{BGetal,JMW}.  The weak current transforms as
$(\bar3_L,1_R)$, and at the zero recoil kinematic point there are two operators
that are relevant for $P^{(Q)} \to V$ transition matrix elements (where
$P^{(b)}=B$, $P^{(c)}=D$, and $V$ is one of the lowest lying vector mesons
$\rho, \omega, K^*, \phi$).  Demanding that the Zweig suppressed
$D_s\to\omega\,\bar\ell\,\nu_\ell$ process vanishes relates the two operators,
yielding \cite{Hooman}
\begin{equation}\label{current}
\bar q_a\, \gamma_\mu (1-\gamma_5)\, Q = 
  \beta\, {\rm Tr} [ {N\!\!\!\!\slash}_{cb}^{\,\dag}
  \gamma_\mu (1-\gamma_5) H_c^{(Q)} \xi_{ba}^\dag ] \,.
\end{equation}
Here repeated $SU(3)$ indices are summed and the trace is over Lorentz indices.
$H^{(Q)}$ contains the ground state heavy meson doublet, $N$ is the nonet
vector meson matrix \cite{JMW}, and $\beta$ is a constant.  The leading
contribution to $R(1)-1$ arises from the Feynman diagrams in Fig.~1.  Diagrams
with a virtual kaon cancel in the double ratio $R$.  Neglecting the vector
meson widths,\footnote{The only significant width is that of the $\rho$ meson. 
Since it occurs in the loop graph involving an $\eta$, neglecting the $\rho$
width amounts to treating $\Gamma_\rho/2m_\eta \ll 1$, which is a reasonable
approximation.} these diagrams yield
\begin{equation}\label{R1}
R(1) - 1 = -\frac{g\,g_2}{12\,\pi^2\,f^2}\, \Big[ 
  G(m_\pi,\Delta^{(b)}) - G(m_\eta,\Delta^{(b)}) - 
  G(m_\pi,\Delta^{(c)}) + G(m_\eta,\Delta^{(c)}) \Big] \,,
\end{equation}
where $\Delta^{(b)}=m_{B^*}-m_B$, $\Delta^{(c)}=m_{D^*}-m_D$, and for 
$m\ge\Delta$,
\begin{equation}\label{GmD}
G(m,\Delta) = \frac{\pi\,m^3}{2\,\Delta} - {(m^2-\Delta^2)^{3/2}\over\Delta}\, 
  \arctan \left( {\sqrt{m^2-\Delta^2}\over \Delta}\right) - \Delta^2 \ln m\,.
\end{equation}
Here $g_2$ is the $\rho\,\omega\,\pi$ coupling, $g$ is the $DD^*\pi$ coupling,
and $f\simeq131\,$MeV is the pion decay constant.  In the nonrelativistic
constituent quark model $g=g_2=1$ \cite{BGetal}, while in the chiral quark
model \cite{chqm} $g=g_2=0.75$.  Experimental data on
$\tau\to\omega\,\pi\,\nu_\tau$ in the region of low $\omega\,\pi$ invariant
mass gives $g_2\simeq0.6$ \cite{dw}.

\begin{figure}[tb]  
\centerline{\epsfysize=2.5truecm \epsfbox{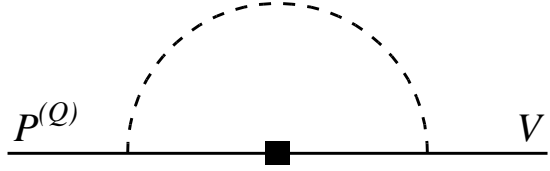}}
\caption[1]{Feynman diagram that gives the leading contribution to $R(1)-1$.
The dashed line is a $\pi$ or an $\eta$.
The black square indicates insertion of the weak current.}
\end{figure}

For small $\Delta$, Eq.~(\ref{R1}) for $R(1)-1$ has a non-analytic $\sqrt{m_q}$
dependence on the light quark masses.  This cannot arise from corrections to
the current in Eq.~(\ref{current}) or to the chiral Lagrangian, and must come
from 1-loop diagrams involving the pseudo-Goldstone bosons $\pi,\,K,\,\eta$. 
Using the measured values of the pion and eta masses, Eqs.~(\ref{R1}) and
(\ref{GmD}) imply $R(1)=1-0.035\,g\,g_2$.  There may be significant corrections
from higher orders in chiral perturbation theory.  However, the smallness of
our result lends support to the expectation that $R(1)-1$ is very close to
zero.  There is no reason to expect any different conclusion over the kinematic
range $1<y<1.5$.

\medskip\noindent
{\bf (ii) Long distance effects and the extraction of $S^{(B\to K^*)}$ 
from $B\to K^*\ell\,\bar\ell$ }
\medskip

The decay rate for $B \to K^*\,\nu_\ell\,\bar\nu_\ell$ could determine
$S^{(B\to K^*)}$ free of theoretical uncertainties.  However, experimental
study of this decay is very challenging.  A more practical approach to
extracting this quantity is to use $B \to K^*\ell\,\bar\ell$.  The differential
decay rate is
\begin{eqnarray}\label{Rrate2}
{{\rm d}\Gamma(B\to K^*\ell\,\bar\ell)\over{\rm d}y} &=& 
  {G_F^2\, |V_{ts}^*V_{tb}|^2 \over 24\,\pi^3} 
  \left({\alpha\over4\pi}\right)^2 m_B\,m_{K^*}^2\, 
  \Big[|\widetilde C_9(y)|^2 + |C_{10}|^2\Big]\, [1+\Delta(y)] \nonumber\\*
&&\phantom{} \times S^{(B\to K^*)}(y)\, [1+d(y)] \,, 
\end{eqnarray}
This and Eq.~(\ref{magic}) allow us to rewrite Eq.~(\ref{SLrate}) as 
\begin{eqnarray}\label{rewrite}
{{\rm d}\Gamma(B\to\rho\,\ell\,\bar\nu_\ell) \over {\rm d}y }
&=& {|V_{ub}|^2\over|V_{ts}^*V_{tb}|^2}\, {8\,\pi^2\over\alpha^2}\, 
  {1\over |\widetilde C_9(y)|^2 + |C_{10}|^2}\, 
  \frac1{1+\Delta(y)}\, \frac1{1+d(y)} \nonumber\\*
&&\phantom{} \times { m_\rho^2 \over m_{K^*}^2}\, 
  \bigg({m_B-m_\rho\over m_B-m_{K^*}}\bigg)^2
  \left|{f^{(D\to\rho)}(y)\over f^{(D\to K^*)}(y)}\right|^2 
  {{\rm d}\Gamma(B\to K^*\ell\,\bar\ell) \over {\rm d}y } \,,
\end{eqnarray}
which can be directly used to extract $|V_{ub}|$.  Unitarity of the CKM matrix
implies that $|V_{ts}^* V_{tb}| \simeq |V_{cs}^* V_{cb}|$ with less than a 3\%
uncertainty.  The fine structure constant, $\alpha=1/129$, is evaluated at the
$W$-boson mass.  $d(y)$ parameterizes long distance effects, and will be
discussed below.  $\Delta(y)$ takes into account the contribution of the
magnetic moment operator, $O_7=(e/16\pi^2)\,m_b\,(\bar
s_L\,\sigma_{\mu\nu}\,b_R)\, F^{\mu\nu}$ (a factor of $-4 G_FV_{ts}^*
V_{tb}/\sqrt{2}$ has been extracted out in the definition of operator
coefficients).  Ref.~\cite{lw} (see also Ref.~\cite{SaYa}) found using heavy
quark symmetry that $\Delta(y) \simeq -0.14-0.08(y-1)$ in the region $1<y<1.5$.
Corrections to this are expected to be small since there are no $1/m_c$
corrections to $\Delta(1)$.  $C_{10}$ is the Wilson coefficient of the operator
$O_{10} = (e^2/16\pi^2)\, (\bar s_L\,\gamma_\mu\,b_L)
(\bar\ell\,\gamma^\mu\gamma_5\,\ell)$.  $\widetilde C_9(y)$ takes into account
the contribution of the four-quark operators, $O_1-O_6$, and the operator $O_9
= (e^2/16\pi^2)\, (\bar s_L\,\gamma_\mu\,b_L)\, (\bar\ell\,\gamma^\mu\,\ell)$. 
In perturbation theory using the next-to-leading logarithmic approximation
\cite{BuMu}
\begin{eqnarray}\label{C9eff}
\widetilde C_9(y) &=& C_9 + h(z,y)\, (3C_1+C_2+3C_3+C_4+3C_5+C_6)
  -\frac12\,h(0,y)\, (C_3+3C_4)  \nonumber\\*
&&\phantom{} -\frac12\,h(1,y)\, (4C_3+4C_4+3C_5+C_6) 
  + \frac29\, (3C_3+C_4+3C_5+C_6) \,,
\end{eqnarray}
where $z=m_c/m_b$.  Here 
\begin{equation}\label{qloops}
h(u,y) = -\frac89\ln u + \frac8{27} + \frac{4}9\,x
  - \frac29\, (2+x) \sqrt{|1-x|} \cases{ 
  \ln{\displaystyle 1+\sqrt{1-x}\over\displaystyle 1-\sqrt{1-x}} - i\pi \,;  
    &  $x<1$ \cr
  2\arctan(1/\sqrt{x-1}) \,; &  $x>1$\,, \cr} \nonumber
\end{equation}
where $x \equiv 4u^2m_b^2/(m_B^2+m_{K^*}^2-2m_B\,m_{K^*}\,y)$.  Using
$m_t=175\,$GeV, $m_b=4.8\,$GeV, $m_c=1.4\,$GeV, $\alpha_s(m_W)=0.12$, and
$\alpha_s(m_b)=0.22$, the numerical values of the Wilson coefficients are
$C_1=-0.26$, $C_2=1.11$, $C_3=0.01$, $C_4=-0.03$, $C_5=0.008$, $C_6=-0.03$,
$C_7=-0.32$, $C_9=4.26$, and $C_{10}=-4.62$.  Of these, $C_9$ and $C_{10}$ are
sensitive to $m_t$ (quadratically for $m_t\gg m_W$).

In Eq.~(\ref{C9eff}) the second term on the right-hand-side, proportional to
$h(z,y)$ comes from charm quark loops.  Since the kinematic region we are
interested in is close to $q^2=4m_c^2$, a perturbative calculation of the
$c\,\bar c$ loop cannot be trusted.  Threshold effects which spoil local
duality are important.  It is these long distance effects that give rise to the
major theoretical uncertainty in the extraction of $|V_{ub}|$ from the $B\to
K^*\ell\,\bar\ell$ differential decay rate using
Eq.~(\ref{rewrite}).\footnote{The four-quark operators involving light $u$,
$d$, and $s$ quarks also have uncertainty from long distance physics.  However,
this is expected to have a very small effect on the $B\to K^*\ell\,\bar\ell$
rate.}  The influence of this long distance physics on the differential decay
rate is parameterized by $d(y)$ in Eq.~(\ref{Rrate2}), where setting $d(y)=0$
gives the perturbative result.

For the part of the $c\,\bar c$ loop where the charm quarks are not far
off-shell, a model for $h(z,y)$ which sums over $1^{--}$ $c\,\bar c$ resonances
is more appropriate than the perturbative calculation.  Consequently, we model
the part of $h(z,y)$ with explicit $q^2$-dependence in Eq.~(\ref{qloops}) with
a sum over resonances \cite{longd} calculated using factorization
\begin{equation}\label{model}
h(z,y) \to -\frac89 \ln{z} + \frac8{27} - \frac{3\pi\kappa}{\alpha^2}\,
  \sum_n { \Gamma_{\psi^{(n)}}\, {\cal B}(\psi^{(n)}\to\ell\,\bar\ell) \over 
  (q^2-M_{\psi^{(n)}}^2)/M_{\psi^{(n)}} + i\Gamma_{\psi^{(n)}}} \,.
\end{equation}
The resonances $\psi^{(n)}$ have masses $3.097\,$GeV, $3.686\,$GeV,
$3.770\,$GeV, $4.040\,$GeV, $4.160\,$GeV, and $4.415\,$GeV, respectively, and
their widths $\Gamma_{\psi^{(n)}}$ and leptonic branching ratios ${\cal
B}(\psi^{(n)}\to\ell\,\bar\ell)$ are known \cite{PDG}.  The factor $\kappa=2.3$
takes into account the deviation of the factorization model \cite{NeSt}
parameter $a_2$ from its perturbative value.  Denoting the value of $\widetilde
C_9(y)$ in this model by $\widetilde C_9'(y)$, its influence on the
differential decay rate is given by $d(y)$ defined as
\begin{equation}\label{dydef}
|\widetilde C_9'(y)|^2+|C_{10}|^2 =
  (|\widetilde C_9(y)|^2+|C_{10}|^2)\, [1+d(y)] \,.
\end{equation}
$d(y)$ is plotted in Fig.~2 (solid curve).  The physical interpretation of the
$1^{--}$ resonances above $4\,$GeV is not completely clear.  It might be more
appropriate to treat them as $D\bar D$ resonances than as $c\,\bar c$ states. 
It is possible that for these resonances factorization as modeled by
Eq.~(\ref{model}) with $\kappa=2.3$ is not a good approximation.  Including
only the first three $1^{--}$ resonances in Eq.~(\ref{model}), yields $d(y)$
plotted with the dashed curve in Fig.~2.  

\begin{figure}[tb]  
\centerline{\epsfysize=8truecm \epsfbox{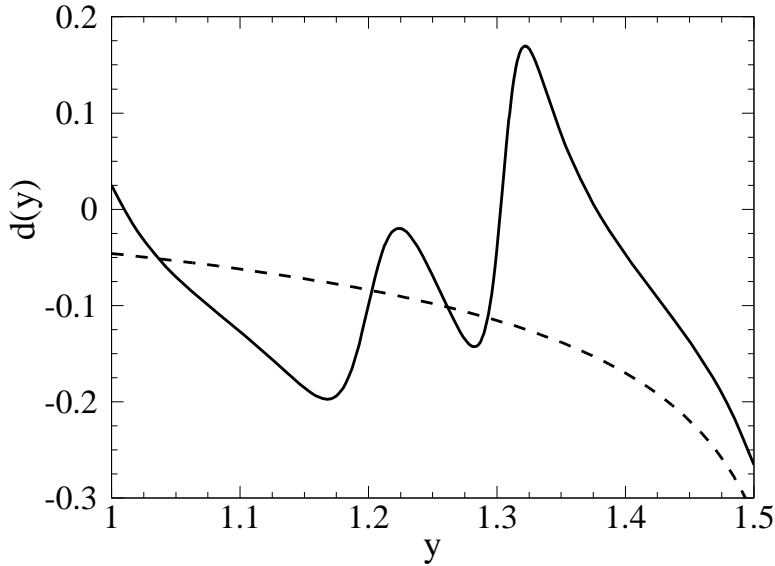}}
\caption[2]{$d(y)$ defined in Eq.~(\ref{dydef}).  The solid curve takes into
account all six $1^{--}$ $c\,\bar c$ resonances according to 
Eq.~(\ref{model}), whereas the dashed curve is obtained including only the 
three lightest ones.}
\end{figure}

This estimate of $d(y)$ based on factorization and resonance saturation differs
from that in Ref.~\cite{lw} in two respects.  Firstly, the phase of $\kappa$ is
viewed as fixed because recent data has determined the sign of the ratio of
factorization model parameters, $a_2/a_1$, and the phase of $a_1$ is expected
to be near its perturbative value \cite{BHP}.  Secondly, since the resonance
saturation model only represents the $c\,\bar c$ loop for charm quarks that are
not far off-shell, we have only used it for the part of $h(z,y)$ in
Eq.~(\ref{qloops}) with explicit $q^2$ dependence, retaining the perturbative
expression for the first two terms, $-(8/9)\ln z+8/27$.  The $\ln z$ term has
dependence on $m_b$, which is clearly short distance in origin.  This reduces
somewhat the magnitude of $d(y)$ and makes it more symmetric about zero
(compare Fig.~2 with Fig.~6 of Ref.~\cite{lw}).  It would be interesting to
have a more physical separation between the long and short distance parts of
the amplitude.

Whether it is reasonable to use factorization for the resonances above $4\,$GeV
can be tested experimentally, since these states cause a very distinctive
pattern in ${\rm d}\Gamma/{\rm d}y$.  In Fig.~3 the shape of ${\rm
d}\Gamma/{\rm d}y$ is plotted in the region $1<y<1.5$ using the resonance
saturation model for $d(y)$ (solid curve).  Experimental support for this shape
would provide evidence that this model correctly describes the long distance
physics parameterized by $d(y)$.  Although $d(y)$ gets as large as $\pm0.2$,
since it oscillates, its influence on the $B\to K^*\ell\,\bar\ell$ decay rate
in the region $1<y<1.5$ is about $-8$\% compared to the perturbative result
(which is plotted with the dotted curve in Fig.~3).  Even if our estimates of
this long distance physics based on factorization and resonance saturation has
a 100\% uncertainty (a prospect that we do not consider particularly unlikely),
it will only cause about a 4\% uncertainty in this determination of $|V_{ub}|$.
Including only the first three $1^{--}$ resonances in the sum in
Eq.~(\ref{model}) yields the dashed curve in Fig.~3.  In this case $d(y)$
causes a $-13$\% change in the $B\to K^*\ell\,\bar\ell$ decay rate in the
region $1<y<1.5$.

\begin{figure}[tb]  
\centerline{\epsfysize=8truecm \epsfbox{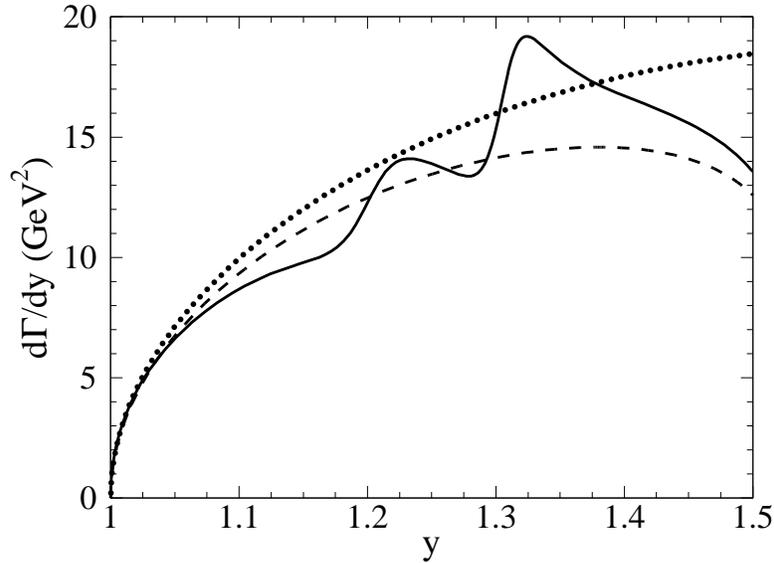}}
\caption[3]{${\rm d}\Gamma(B\to K^*\ell\,\bar\ell)/{\rm d}y$ in units of 
$[|\widetilde C_9(1)|^2+|C_{10}|^2]\, m_Bm_{K^*}^2\, 
G_F^2|V_{ts}^*V_{tb}|^2\alpha^2/(384\pi^5)$ as given in Eq.~(\ref{Rrate2}).  
The solid curve takes into account all six $1^{--}$ $c\,\bar c$ resonances, 
the dashed curve includes only the three lightest ones, and the dotted curve 
is the perturbative result (i.e., $d(y)=0$).}
\end{figure}

\medskip\noindent
{\bf (iii) Nearer term prospects}
\medskip

Without information on the $y$ spectrum for the $B$ decay rates in
Eq.~(\ref{rewrite}), it is still possible to determine $|V_{ub}|$ by comparing
the branching ratios for $B\to \rho\,\ell\,\bar\nu_\ell$ and $B\to
K^*\ell\,\bar\ell$ in the region $1<y<1.5$.  Integrating Eq.~(\ref{rewrite})
over $1<y<1.5$ we can write
\begin{eqnarray}\label{crts}
\Gamma(\bar B^0\to \rho^+\,\ell\,\bar\nu_\ell) \Bigl|_{y<1.5} &=&
  {|V_{ub}|^2\over|V_{ts}^*V_{tb}|^2}\, {8\pi^2\over\alpha^2}\,
  {1\over \overline C{}_9^2+C_{10}^2}\,
  \frac1{(1+\overline\Delta)}\, \frac1{(1+\overline d)} \\*
&&\phantom{} \times \frac{m_\rho^2}{m_{K^*}^2}\, 
  \bigg({m_B-m_\rho\over m_B-m_{K^*}}\bigg)^2\,
  \left|{f^{(D\to\rho)}(1)\over f^{(D\to K^*)}(1)}\right|^2
  \Gamma(B\to K^*\ell\,\bar\ell)\Big|_{y<1.5} \nonumber\,. 
\end{eqnarray}
Here the barred quantities, $\overline C{}_9^2$, $\overline\Delta$, and
$\overline d$ denote the averages of $|\widetilde C_9(y)|^2$, $\Delta(y)$, and
$d(y)$ weighted with $S^{(B\to K^*)}(y)$.  Using the shape for $S^{(B\to K^*)}$
predicted from heavy quark symmetry, we find $\overline C_9=4.58$,
$\overline\Delta=-0.16$, and $\overline d=-0.08$.  Note that the $y$-dependence
of $\widetilde C_9$ is small and $\overline C_9$ is close to $C_9$.  In
Eq.~(\ref{crts}) the $y$-dependence of the ratio $f^{(D\to\rho)}(y)/f^{(D\to
K^*)}(y)$ has been neglected.  If the shape of these form factors can be
approximated with a pole form, then the pole masses of $2.56\,{\rm GeV}$ for
$f^{(D\to K^*)}$ and $2.45\,{\rm GeV}$ for $f^{(D\to\rho)}$ (corresponding to
$D^{**}_s$ and to $D^{**}$) imply that $|f^{(D\to\rho)}(y)/f^{(D\to
K^*)}(y)|^2$ varies by less than 1\% over the range $1<y<1.5$.  $SU(3)$
symmetry and the measured $D\to K^*$ form factors imply that
$\delta^{(D\to\rho)}$ contributes only about 23\% of the
$D\to\rho\,\bar\ell\,\nu_\ell$ decay rate.  Using this prediction for
$\delta^{(D\to\rho)}$, and assuming that $f^{(D\to\rho)}$ and $f^{(D\to K^*)}$
have the same $y$-dependence, yields ${\cal
B}(D\to\rho^0\,\bar\ell\,\nu_\ell)/{\cal B}(D\to\bar
K^{*0}\,\bar\ell\,\nu_\ell)=0.044\,|f^{(D\to\rho)}(1)/f^{(D\to K^*)}(1)|^2$.

In the region $q^2=(p_\ell+p_{\bar\ell})^2<m_{J/\psi}^2$ (corresponding roughly
to $y>2$), one cannot use the double ratio and Eq.~(\ref{rewrite}).  Moreover,
the $O_7$ contribution to the $B\to K^*\ell\,\bar\ell$ rate is large and
proportional to $1/q^2$, so the (leading order) heavy quark symmetry relations
between the tensor and (axial-)vector form factors\footnote{It was argued in
Ref.~\cite{largew} that heavy quark symmetry can be used even at small $q^2$.}
introduce a significant uncertainty.  For $q^2<m_{J/\psi}^2$, one can do better
using $SU(3)$ flavor symmetry alone to predict ${\rm
d}\Gamma(B\to\pi\,\ell\,\bar\nu_\ell)/{\rm d}q^2$ from a measurement of ${\rm
d}\Gamma(B\to K\ell\,\bar\ell)/{\rm d}q^2$.  Since this region is far from
$q^2_{\rm max}$, the $B^*$ pole contribution \cite{Bpi} is unlikely to upset
the $SU(3)$ relations.  The $O_7$ contribution to ${\rm d}\Gamma(B\to
K\ell\,\bar\ell)/{\rm d}q^2$ is at the $10-15$\% level, fairly independent of
$q^2$.  In the region $(1-2){\rm GeV}^2<q^2<m_{J/\psi}^2$, neglecting
$m_{K,\pi}^2/m_B^2$,
\begin{equation}\label{Kpi}
{ {\rm d}\Gamma(\bar B^0 \to \pi^+\,\ell\,\bar\nu_\ell) \over {\rm d}q^2 } =
  {|V_{ub}|^2\over|V_{ts}^*V_{tb}|^2}\, {8\pi^2\over\alpha^2}\,
  {1\over |\widetilde C_9(q^2)+2C_7|^2+|C_{10}|^2}\,
{ {\rm d}\Gamma(B\to K\,\ell\,\bar\ell) \over {\rm d}q^2 } \,. 
\end{equation}
A similar relation also holds for integrated rates.  

A measurement of the $B\to K^*\ell\,\bar\ell$ decay rate is unlikely before the
Tevatron run II.  Without this measurement, one has to rely on predicting the
$B\to\rho$ form factors from $D\to\rho$ using heavy quark symmetry, or from
$D\to K^*$ using both chiral and heavy quark symmetries.  As discussed
following Eq.~(\ref{magic}), recent experimental data \cite{E791a} suggests
that the $SU(3)$ relation between $f^{(D\to K^*)}$ and $f^{(D\to\rho)}$ is not
violated by more than 15\%.  Heavy quark symmetry and the measured $D\to K^*$
form factors in Eq.~(\ref{ffexp}) imply that the $\bar
B^0\to\rho^+\ell\,\bar\nu_\ell$ branching ratio in the region $1<y<1.5$ is
$5.9\,|V_{ub}|^2$.  The measured decay rate ${\cal B}(\bar
B^0\to\rho^+\ell\,\bar\nu_\ell)=(2.5\pm0.4^{+0.5}_{-0.7}\pm0.5)\times10^{-4}$
\cite{excl} together with $|V_{ub}|\sim0.003$ imply that about 20\% of $\bar
B^0\to\rho^+\ell\,\bar\nu_\ell$ decays are in the range $1<y<1.5$.

Despite the presence of long distance effects associated with the $c\,\bar c$
resonance region, the $B\to K^*\ell\,\bar\ell$ rate can be used in
Eq.~(\ref{rewrite}) to determine $|V_{ub}|$ with a theoretical uncertainty that
is about 10\%.  Experimental verification of the distinctive $y$-dependence of
the differential rate associated with the $1^{--}$ resonances above $4\,$GeV
(see Fig.~3) would reduce the theoretical uncertainty from long distance
effects.  A precise value of $|V_{ub}|$ may be available from other processes,
e.g., the hadronic invariant mass spectrum in inclusive $\bar B\to
X_u\ell\,\bar\nu_\ell$ decay \cite{FLW} or from lattice QCD results on
exclusive form factors \cite{lattice} before the $B\to K^*\ell\,\bar\ell$ decay
rate is measured.  In that case, Eq.~(\ref{rewrite}) gives an accurate standard
model prediction for the $B\to K^*\ell\,\bar\ell$ decay rate in the region
$1<y<1.5$.  Comparison with data may signal new physics or provide stringent
constraints on extensions of the standard model.

\medskip\noindent {\bf Acknowledgements} \medskip

We thank J. Kroll and S. Stone for discussions.
I.W.S.\ and M.B.W.\ were supported in part by the U.S.\ Dept.\ of Energy under
Grant no.\ DE-FG03-92-ER~40701.  Z.L. was supported in part by the U.S.\ Dept.\
of Energy under grant no.\ DOE-FG03-97ER40506 and by the NSF grant PHY-9457911.

{\tighten

} 

\end{document}